\documentclass[runningheads]{llncs}
\usepackage[T1]{fontenc}
\usepackage{graphicx,verbatim}
\usepackage[utf8]{inputenc}
\usepackage{graphicx}
\usepackage{amssymb}
\usepackage{amsmath}
\usepackage{xcolor}
\usepackage{subcaption}
\usepackage{hyperref}
\usepackage{multirow}
\usepackage{booktabs}
\begin{document}
\title{SurgRFO: Foundation Model Based Compositional Synthesis of Critical Retained Foreign Objects in Intraoperative Chest X-rays}
%

\author{
Yuanyun Hu\inst{1,2}\textsuperscript{\dag} \and
Yuli Wang\inst{1,3}\textsuperscript{\dag} \and
Noemi Acevedo Rodriguez\inst{3} \and
Ronald Yang\inst{3} \and
Wen-Chi Hsu\inst{4} \and
Siwei Luo\inst{5} \and
Zihao Bai\inst{2} \and
Jing Wu\inst{6} \and
Yuwei Dai\inst{3} \and
Shaoju Wu\inst{3} \and
Jonathon Lindquist\inst{3} \and
Justin Honce\inst{3} \and
Premal Trivedi\inst{3} \and
Zhicheng Jiao\inst{7} \and
Ihab Kamel\inst{3} \and
Elliott Haut\inst{1} \and
Pamela Johnson\inst{1} \and
John Eng\inst{1} \and
Cheng Ting Lin\inst{1} \and
Nan Su\inst{2} \and
Bo Chen\inst{8}\textsuperscript{*} \and
Sun Yu\inst{9} \and
Harrison Bai\inst{3}\textsuperscript{*}
}

\authorrunning{Hu et al.}

\institute{
Department of Radiology and Radiological Science, Johns Hopkins University School of Medicine, Baltimore, USA \and
Department of Electronic Engineering, Tsinghua University, China \and
Department of Radiology, University of Colorado Denver Anschutz Medical Campus, USA \and
Department of Medical Imaging and Intervention, Chang Gung Memorial Hospital at Linkou, Taiwan \and
Department of Electronic Engineering, Southwest University of Science and Technology, China \and
Department of Radiology, Second Xiangya Hospital, Central South University, Changsha, Hunan, China \and
Department of Diagnostic Imaging, Brown University Health, Providence, USA \and
School of Clinical Medicine, Tsinghua University, China \and
Department of Electrical and Computer Engineering, Johns Hopkins University, Baltimore, USA
}

\begingroup
\renewcommand\thefootnote{\dag}
\footnotetext{These two authors contributed equally to this work.}
\renewcommand\thefootnote{*}
\footnotetext{Corresponding author: \texttt{cba04570@btch.edu.cn} and \texttt{harrison.bai@cuanschutz.edu}}
\endgroup
  
\maketitle
\begin{abstract}
Critical retained foreign objects (RFOs) on intraoperative chest radiographs are rare but high-risk events. Their scarcity limits robust automated detection model training and generalization. We introduce \textbf{SurgRFO}, a two-stage synthesis framework for generating realistic RFO-present intraoperative chest X-rays. In \textbf{Stage~1}, a Roentgen chest X-ray foundation model is fine-tuned on surgical-domain images to generate realistic RFO-free backgrounds that preserve anatomy, indwelling lines and tubes, and intraoperative imaging characteristics. In \textbf{Stage~2}, a lightweight generator trained on localized RFO patches from limited positive cases synthesizes diverse RFO instances, which are composited onto generated backgrounds using conditional Poisson fusion to improve photometric consistency. We evaluate SurgRFO through (i) a blinded clinician study assessing realism and clinical plausibility, and (ii) downstream detection experiments in which synthesized data are used to augment Faster R-CNN, YOLOv8, and RetinaNet. SurgRFO consistently improves sensitivity at low false-positive-per-image (FPPI) operating points on internal and external test sets. Clinician ratings indicate that the synthesized images achieve realism comparable to real intraoperative images. Ablation analyses further examine fusion strategies and synthesis scale. Ethical safeguards for synthetic surgical data are also discussed. The code is publicly available on GitHub at \url{https://anonymous.4open.science/r/SurgRFO-7CB5/}.
\end{abstract}

%
\section{Introduction}

Critical retained foreign objects (RFOs), such as surgical sponges and needle fragments, are rare but high-impact patient safety events. Unlike non-critical, intentionally placed objects, critical RFOs are unintended remnants that may cause severe complications, re-operations, and substantial medico-legal and financial burden.~\cite {rigamonti2025retained,kawakubo2023deep,regenbogen2009prevention,moffatt2014risk,rajagopal2002gossypiboma,stawicki2009retained,sun2007gossypiboma}. Clinically, intraoperative imaging aims to rule out RFOs before procedure completion, where missed detections can have serious consequences.
However, work explicitly targeting critical RFOs remains limited. Most prior studies focus on standardized chest X-rays (CXRs) and predominantly address \emph{non-critical} objects~\cite{xue2015foreign,viola2001rapid,santosh2021generic,kufel2023chest}. Robust detection of critical RFOs in real-world intraoperative X-rays, therefore, remains an open and clinically significant challenge.

Critical RFO detection relies on intraoperative radiographs acquired under far less controlled conditions than standard CXRs, with cluttered scenes, overlapping instruments, variable anatomy exposure, and heterogeneous imaging settings.
Two major challenges arise. First, visual cues are often subtle and confounded by surgical context. Sponges or needle fragments may be blood-soaked, partially occluded, or blended with surrounding tissue, producing weak signals that are difficult to recognize, even for experienced clinicians.~\cite{kumar2017imaging,o2003imaging}. False negatives are therefore not uncommon, and prior studies suggest that intraoperative imaging may fail to detect a substantial fraction of RFO cases~\cite{cima2008incidence}.
Second, positive cases are extremely rare and difficult to scale. Even at a large hospital, only low hundreds of cases may accumulate over decades, and no public dataset dedicated to critical RFOs exists.
Together, these constraints create a fundamental limitation: accurate detection requires sensitivity to subtle local patterns, yet available data provide limited exposure to such events, challenging conventional supervised learning approaches.

\begin{figure}[tbp]
\includegraphics[width=0.98\textwidth]{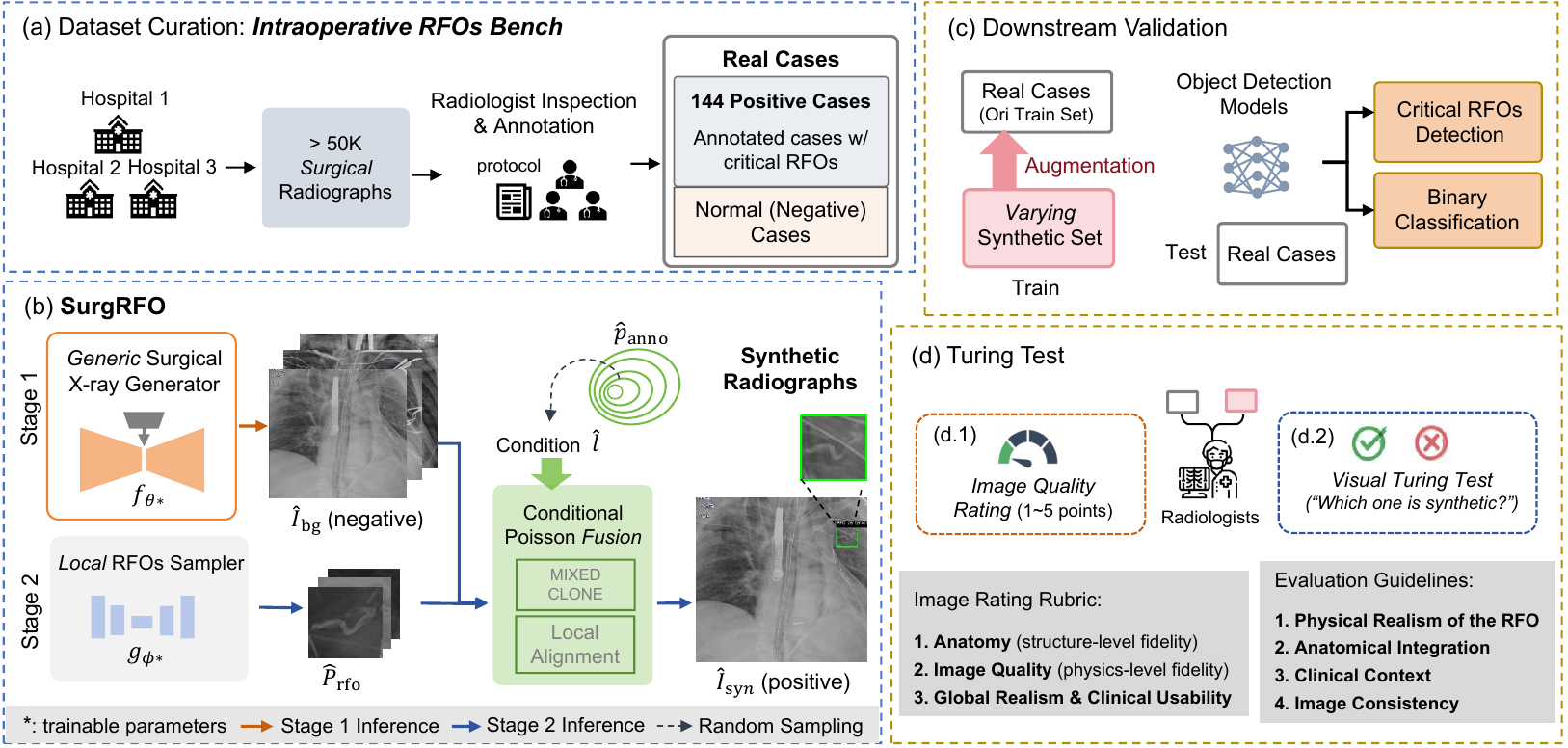}
\centering
\caption{
Overview of the \textbf{Intraoperative RFO Bench} and \textbf{SurgRFO} pipeline. (a) Multi-institutional intraoperative radiographs are curated and expert-annotated as real positive (critical RFO) and negative cases. (b) SurgRFO applies a two-stage synthesis framework, combining surgical X-ray background generation with localized RFO sampling and fusion to create annotated synthetic images. (c) Synthetic data are used for training augmentation and evaluated on held-out real cases for critical RFO detection and classification. (d) A radiologist-led Turing test assesses visual realism through confidence-based quality ratings and image discrimination tasks.
} 
\label{fig_overview}
\end{figure}

Synthetic data augmentation is a natural strategy to mitigate data scarcity.
Recent advances in diffusion models (DMs), which generate images through iterative denoising of a learned reverse process~\cite{ho2020denoising}, have enabled high-fidelity and diverse X-ray synthesis~\cite{bluethgen2025vision,weber2023cascaded,xie2025sv,wang2025dataset,zhou2025dataset}, supporting augmentation for imbalanced medical imaging tasks~\cite{khosravi2024synthetically,prakash2025evaluating,huijben2024denoising,mahaulpatha2024ddpm,hsu2025mri,wu2025vision}.
However, fully end-to-end generation of critical RFO radiographs remains unreliable. The same rarity and subtle visual signatures that hinder detection also limit generative fidelity: standard pipelines struggle to simultaneously preserve complex surgical context and localized RFO cues, often leading to visually implausible samples for training detectors (see comparisons in Section~\ref{sec:Results}).
To address this gap, we propose \textbf{SurgRFO}, a two-stage synthesis framework for generating surgical radiographs with critical RFOs.
Stage~1 learns a strong prior over cluttered surgical radiographs by training a latent diffusion model (LDM) as a generic surgical X-ray generator.
Stage~2 trains a lightweight U-Net to model local critical RFO appearance, then conditionally fuses the sampled RFO content into Stage~1 backgrounds via a Poisson-based blending mechanism guided by a prior over plausible RFO locations and sizes.

We also release \textbf{Intraoperative RFO Bench}, the first publicly available dataset dedicated to \emph{critical} RFOs, comprising 144 expert-annotated positive cases curated from 18 years of surgical radiographs from two health systems (Site 1 and Site 2).
Augmenting training with SurgRFO yields consistent improvements across multiple detector families. In a radiologist-led Turing-style evaluation, synthetic images are frequently indistinguishable from real cases. Together, these results demonstrate the value of releasing this benchmark and the effectiveness of the proposed framework for advancing critical RFO detection.

\section{Method}

\subsection{Dataset Curation}\label{dataset}

Under institutional review board approval, we retrospectively curated surgical chest radiographs from the Site 1 Health System, comprising 144 \emph{critical} RFO-positive cases and 944 negative cases with no RFOs. An additional 20 \emph{critical} RFO-positive cases were collected from Site 2 for external evaluation. Positives were identified via structured report search and expert verification using radiology reports and operative notes. All images were de-identified and annotated by trained radiologists under a standardized protocol, including image-level labels and object-level annotations (bounding boxes or polygons), with each foreign object categorized as critical or non-critical.
Patient-level splits into training, validation, and test sets were applied consistently across cohorts to prevent information leakage. Training images were preprocessed with center cropping, resolution normalization, and artifact filtering.

\subsection{SurgRFO}
Let $\Omega\subset\mathbb{R}^2$ denote the image lattice, and let $I\in\mathbb{R}^{H\times W}$ denote a radiograph defined on $\Omega$. Two datasets are considered,
$\mathcal{D}_{\mathrm{bg}}=\{I_i\}_{i=1}^{N_{\mathrm{bg}}}$ and
$\mathcal{D}_{\mathrm{rfo}}=\{(I_j,M_j,y_j)\}_{j=1}^{N_{\mathrm{rfo}}}$.
The set $\mathcal{D}_{\mathrm{bg}}$ contains surgical-domain negative radiographs that provide RFO-free backgrounds. The set $\mathcal{D}_{\mathrm{rfo}}$ contains radiographs with annotated critical RFOs. The mask $M_j\subset\Omega$ specifies the RFO support, and the label $y_j\in\mathcal{Y}$ specifies the semantic type.

SurgRFO is a two-stage synthesis framework that decouples global background modeling from local critical RFO modeling. Stage~1 learns a generic prior over surgical radiograph backgrounds.
Stage~2 learns a local sampler for critical RFO appearance and composes sampled RFO evidence onto Stage~1 backgrounds through a controlled fusion mechanism.

\subsubsection{Stage~1: Generic Surgical Radiograph Generator.}
A background generator is defined as a conditional LDM $f_{\theta}$ parameterized by
$\theta$. Background sampling is written as
\begin{equation}
\hat I_{\mathrm{bg}} \sim p_{\theta}(I \mid t, z_1), \qquad z_1\sim\mathcal{N}(0,I),
\end{equation}
where $\hat I_{\mathrm{bg}}$ denotes an RFO-free surgical background, $t$ denotes a domain anchor text token, and $z_1$ denotes the diffusion noise seed. Stage~1 is designed for domain-level background modeling rather than instance-specific textual control. A fixed domain anchor prompt is therefore used, and diversity is induced primarily by the stochasticity of $z_1$. In implementation, we adapt RoentGen~\cite{bluethgen2025vision}, an advanced CXRs LDM, to surgical X-ray domain leveraging $\mathcal{D}_{\mathrm{bg}}$.

\subsubsection{Stage~2: Local RFO Sampler and Conditional Fusion.}
A local RFO sampler is defined by a compact generator $g_{\phi}$ that models patch-level RFO appearance. For each annotated sample $(I, M,y)\in\mathcal{D}_{\mathrm{rfo}}$, an RFO-centered canonical patch is extracted by a mask-driven cropping operator $\mathcal{C}_{M}$,
\begin{equation}
P = \mathcal{C}_{M}(I).
\end{equation}
A plausible RFO patch is sampled from a noise seed $z_2$ as
\begin{equation}
\hat P_{\mathrm{rfo}} \sim p_{\phi}(P \mid z_2), \qquad z_2\sim\mathcal{N}(0,I),
\end{equation}
where $p_{\phi}$ denotes the implicit patch distribution defined by $g_{\phi}$.

An anatomically plausible placement is specified by a location-scale variable
$\ell=(x,y,s)$, where $(x,y)$ denotes the insertion center on the background canvas and $s$ denotes the insertion scale. An empirical placement prior $\hat p_{\mathrm{anno}}(\ell)$ is estimated from annotation statistics, and a concrete placement parameter is drawn as
$\hat\ell \sim \hat p_{\mathrm{anno}}(\ell)$. 
A placement operator $\mathcal{T}_{\hat\ell}$ maps both the sampled patch and its support to the full image grid,
\begin{equation}
\hat P_{\hat\ell}=\mathcal{T}_{\hat\ell}(\hat P_{\mathrm{rfo}}), \qquad
M_{\hat\ell}=\mathcal{T}_{\hat\ell}(S(\hat P_{\mathrm{rfo}})),
\end{equation}
where $S(\cdot)$ produces a coarse support mask in patch coordinates and $M_{\hat\ell}$ denotes the corresponding insertion support on the $H\times W$ canvas.

The final composite image $I_{\mathrm{syn}}$ is defined by an aligned mixed-cloning objective over the insertion region $\Omega_{\hat\ell}=\{\mathbf{u}\in\Omega: M_{\hat\ell}(\mathbf{u})=1\}$,
\begin{equation}
I_{\mathrm{syn}}
=\arg\min_{I}\;
\sum_{\mathbf{u}\in\Omega_{\hat\ell}}
\left\|\nabla I(\mathbf{u}) -
\Big(\beta\,\nabla \mathcal{A}_{\alpha}(\hat P_{\hat\ell})(\mathbf{u})
+(1-\beta)\,\nabla \hat I_{\mathrm{bg}}(\mathbf{u})\Big)\right\|_2^2,
\end{equation}
where $\nabla$ denotes the spatial gradient operator. The parameter $\beta\in[0,1]$ controls the strength of mixed cloning by balancing patch-driven gradients and background-driven gradients.
The operator $\mathcal{A}_{\alpha}$ performs local intensity alignment of the placed patch to the background within $M_{\hat\ell}$. The parameter $\alpha\in[0,1]$ controls the degree of alignment by interpolating between the raw patch intensity and an intensity level matched to the local background statistics.

\begin{figure}[t]
\centering
\includegraphics[width=0.95\linewidth]{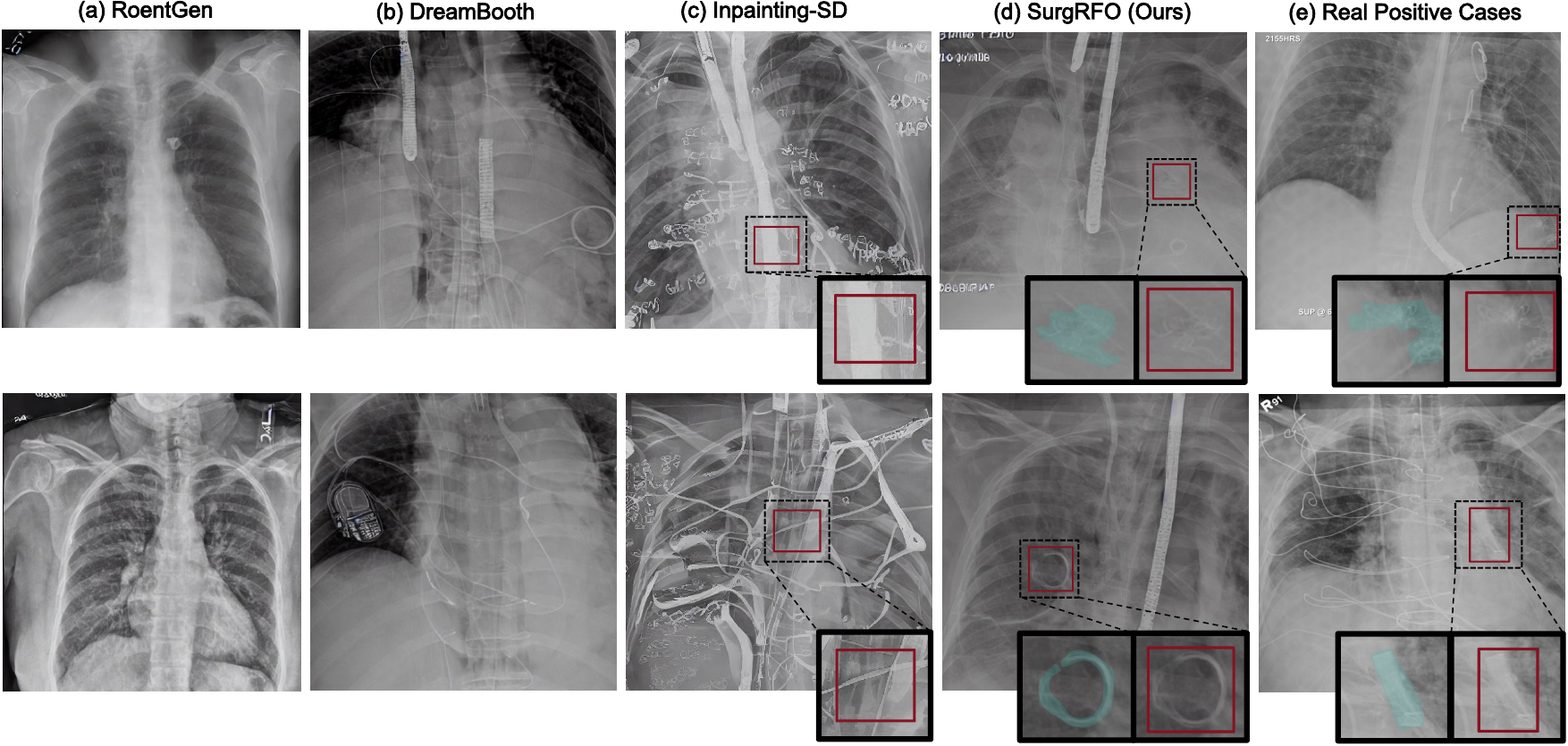}
\caption{\textbf{Visualization of synthesis results.}
Each row shows one representative case. Columns (a-b) show Stage-1 RFO-negative background generation using the original RoentGen and DreamBooth fine-tuning. Columns (c–d) compare Stage-2 RFO-positive synthesis using inpainting Stable Diffusion versus the proposed SurgRFO framework, with real positive cases shown in (e) for reference. Black insets provide zoomed views; red boxes mark target regions, and turquoise overlays indicate critical RFOs. SurgRFO maintains surgical radiograph fidelity while generating clinically plausible critical RFO appearances.}
\label{fig:synthesis_vis}
\end{figure}

\section{Experiments and Results}\label{sec:Results}
\smallskip

\noindent\textbf{Experimental setup.}
SurgRFO is evaluated on the Intraoperative RFOs Bench, which is split into training, validation, and test sets with a 6:1:3 patient-level ratio. In Stage~1, we instantiate the generator from RoentGen under the Stable Diffusion (SD) latent diffusion formulation~\cite{rombach2022high}, and adapt it to the surgical background cohort using a fixed domain prompt and varying noise seeds for diversity. Stage~2 trains a compact U-Net diffusion sampler~\cite{dhariwal2021diffusion} on RFO-centered crops from the positive cohort and synthesizes RFO-positive composites via conditional fusion.
We evaluate SurgRFO from:
(i) synthesis quality, and
(ii) the impact of synthetic augmentation on downstream critical RFO detection.
We use FID~\cite{heusel2017gans} and MS-SSIM~\cite{wang2003multiscale} for qualifying synthesis quality. 
Baselines include DreamBooth fine-tuning~\cite{ruiz2023dreambooth}, and an inpainting SD~\cite{rombach2022high}. RoentGen serves as the common initialization for both the baselines and SurgRFO.
Downstream performance is evaluated on held-out real radiographs using detection metrics, including mean average precision (mAP), false negative rate (FNR), false omission rate (FOR), free-response receiver operating characteristic (FROC), and binary classification accuracy (ACC) for positive versus negative prediction.

\subsection{Synthesis Quality}

\begin{table}[t]
\centering
\scriptsize
\setlength{\tabcolsep}{6pt}
\caption{\textbf{Quantitative assessment of image fidelity and diversity for Stage-1 RFO-negative synthesis.}
FID and MS-SSIM for surgical background generation. \textbf{Bold} indicates the best, and \underline{underlined} indicates the second best among adapted models across training steps. 
}

\vspace{0.5em}
\begin{tabular}{l c c c}
\toprule
\textbf{Method} & \textbf{Steps} & \textbf{FID $\downarrow$} & \textbf{MS-SSIM $\downarrow$} \\
\midrule
Original RoentGen SD (no fine-tuning) & --  & 175.167 & 0.262 $\pm$ 0.130 \\
DreamBooth SD (few-shot fine-tuning)  & --  & 80.963  & 0.424 $\pm$ 0.085 \\
\midrule
\multirow{5}{*}{Ours} & 1k & 48.568 & 0.406 $\pm$ 0.071 \\
                      & 2k & \textbf{43.675} & 0.474 $\pm$ 0.079 \\
                      & 3k & \underline{46.100} & 0.460 $\pm$ 0.052 \\
                      & 4k & 57.896 & \textbf{0.362} $\pm$ \textbf{0.054} \\
                      & 5k & 47.948 & \underline{0.393} $\pm$ \underline{0.059} \\
\bottomrule
\end{tabular}
\label{tab:stage1_fid_msssim}
\end{table}

Table~\ref{tab:stage1_fid_msssim} evaluates Stage-1 RFO-negative synthesis on 5k images. Compared with the original RoentGen SD and a DreamBooth-adapted baseline, SurgRFO achieves consistently lower FID across checkpoints, with the best fidelity at 2k steps (FID 43.675). While DreamBooth improves domain alignment compared to the original RoentGen Model, it remains inferior to SurgRFO. 
Diversity improves with training and is best at 4k steps (lowest MS-SSIM), suggesting that Stage~1 captures surgical radiograph appearance with a selectable fidelity–diversity trade-off along training.
Table~\ref{tab:stage2_fid_msssim} assesses final RFO-positive composites after Stage~2. SurgRFO improves fidelity over both the original RoentGen and inpainting baselines. Because Stage~2 edits only localized regions, MS-SSIM is dominated by preserved backgrounds and is reported as a secondary diagnostic, while FID remains the primary criterion for comparison.
Visual comparisons (Fig.~\ref{fig:synthesis_vis}) further demonstrate that end-to-end baselines struggle to maintain radiographic realism and render subtle RFO cues under limited positive data. In contrast, SurgRFO preserves global surgical fidelity and produces coherent, clinically plausible critical RFO appearances, as evidenced by the zoomed-in insets.

\begin{table}[t]
\centering
\scriptsize
\setlength{\tabcolsep}{10pt}
\caption{\textbf{Quantitative assessment for Stage-2 RFO-positive synthesis.}
\textbf{Bold} indicates the best FID among compared methods.}

\vspace{0.5em}
\begin{tabular}{l c c}
\toprule
\textbf{Method} & \textbf{FID $\downarrow$} & \textbf{MS-SSIM $\downarrow$} \\
\midrule
Original RoentGen SD & 197.837 & 0.262 $\pm$ 0.130 \\
Inpainting SD        & 154.523 & 0.118 $\pm$ 0.045 \\
Ours                 & \textbf{69.178} & 0.480 $\pm$ 0.064 \\
\bottomrule
\end{tabular}
\label{tab:stage2_fid_msssim}
\end{table}

\subsection{Downstream Detection with SurgRFO Augmentation}

\noindent\textbf{Internal evaluation.} The downstream evaluation assesses whether SurgRFO improves critical RFO detection on real radiographs. As shown in Table~\ref{tab:downstream_results}, augmenting training with synthetic positives yields consistent gains across three detector families, suggesting that synthetic samples broaden effective training coverage rather than favoring a specific detector. Faster R-CNN demonstrates marked performance improvement with a substantial reduction in missed positive cases. RetinaNet shows monotonic gains as more synthetic samples are added. YOLOv8l performs poorly when trained on the base regime alone but achieves meaningful detection once synthetic samples are introduced, indicating particular benefit under limited positive supervision.

\noindent\textbf{External evaluation.}
Generalization is further evaluated on an external real-world cohort from the Site 2 hospital system.
After SurgRFO-based augmentation, all models demonstrate clear improvement, with consistent reductions in false negatives and gains in overall accuracy. RetinaNet shows the largest benefit, with mAP@0.3 improving by +0.260, FNR decreasing by -0.750, FOR decreasing by -0.321, FROC increasing by +0.083, and ACC improving by +0.428, transitioning to meaningful retrieval performance. Faster R-CNN and YOLOv8l also exhibit reduced FNR (-0.250 each) and improved accuracy (+0.142 each), with YOLOv8l achieving a notable FROC gain (+0.250). These findings indicate that synthesized positives provide effective supervision beyond the development cohort and enhance robustness under dataset shift.

\begin{table}[t]
\centering
\setlength{\tabcolsep}{4pt}
\caption{\textbf{Downstream critical RFO detection with synthetic augmentation.}
Detectors are trained on the base real set with additional SurgRFO-generated positive samples at varying amounts. \textbf{Bold} marks the best result and \underline{underlined} marks the second best within each detector block.}
\vspace{0.5em}
\scriptsize
\begin{tabular}{llccccc}
\toprule
\textbf{Model} & \textbf{Training Set} &
\textbf{$\mathbf{mAP}@0.3$ $\uparrow$} &
\textbf{FNR $\downarrow$} &
\textbf{FOR $\downarrow$} &
\textbf{FROC $\uparrow$} &
\textbf{ACC $\uparrow$} \\
\midrule
\multirow{4}{*}{Faster R-CNN}
& Base       & 0.184 & 78.79\% & 8.72\% & 0.242 & 0.672 \\
& Base+500   & 0.304 & 66.67\% & 7.48\% & 0.424 & 0.928 \\
& Base+1000  & \underline{0.390} & \underline{60.61\%} & \underline{6.87\%} & \underline{0.455} & \underline{0.931} \\
& Base+2000  & \textbf{0.510} & \textbf{33.33\%} & \textbf{3.93\%} & \textbf{0.636} & \textbf{0.954} \\
\midrule
\multirow{4}{*}{RetinaNet}
& Base       & 0.099 & 72.73\% & 8.11\% & 0.455 & 0.921 \\
& Base+500   & 0.364 & 63.64\% & 7.17\% & 0.485 & 0.931 \\
& Base+1000  & \underline{0.455} & \underline{54.55}\% & \underline{6.23\%} & \underline{0.667} & \underline{0.938} \\
& Base+2000  & \textbf{0.564} & \textbf{36.36\%} & \textbf{4.26\%} & \textbf{0.727} & \textbf{0.954} \\
\midrule
\multirow{4}{*}{YOLOv8l}
& Base       & 0.000 & 100.00\% & 10.82\% & --    & 0.892 \\
& Base+500   & 0.290 & 69.70\% & 7.85\% & 0.303 & 0.918 \\
& Base+1000  & \textbf{0.357} & \textbf{60.61\%} & \textbf{6.87\%} & \underline{0.394} & \textbf{0.931} \\
& Base+2000  & \underline{0.355} & \underline{63.64}\% & \underline{7.24}\% & \textbf{0.394} & \underline{0.921} \\
\bottomrule
\end{tabular}
\label{tab:downstream_results}
\end{table}

\section{Radiologist-driven Turing Test}\label{sec:turing}

To assess clinical plausibility beyond automated metrics, we conducted a blinded radiologist evaluation using a Turing-style protocol (Fig.~\ref{fig_overview} d) comprising two tasks: (1) blinded image quality rating without disclosure of provenance, and (2) explicit discriminability between real and synthetic images.

\smallskip
\noindent\textbf{Turing Test 1: Blinded image quality rating.}
Radiologists independently scored randomized radiographs using a five-point Likert rubric. As shown in Table~\ref{tab:turing} (a), synthetic images received quality ratings not significantly different from real images under Welch’s test, indicating the synthetic cohort preserves clinically plausible radiographic appearance under blinded assessment.

\begin{table}[t]
\centering
\caption{Radiologist-driven Turing tests for clinical realism.}
\label{tab:turing}
\scriptsize
\setlength{\tabcolsep}{3pt}
\renewcommand{\arraystretch}{1}

\begin{minipage}[t]{0.485\textwidth}
\centering
\textbf{(a) Blinded image-quality ratings on a 1--5 Likert scale.} Values are mean$\pm$standard deviation. Welch's test compares real and synthetic images per rater.

\begin{tabular}{lccc}
\toprule
 & Real & Synthetic & Welch $p$\\
\midrule
Rater 1 & 2.86$\pm$1.11 & 2.47$\pm$1.02 & 0.166 \\
Rater 2 & 2.64$\pm$1.22 & 2.31$\pm$0.74 & 0.220\\
\bottomrule
\end{tabular}
\end{minipage}
\hfill
\begin{minipage}[t]{0.485\textwidth}
\centering
\textbf{(b) Discrimination between real and synthetic images.} Metrics include accuracy, F1 score, the fraction of synthetic images labeled as real, and confidence--accuracy AUC.\\

\begin{tabular}{lcccc}
\toprule
 & Acc & F1 & Syn$\to$Real & ConfAUC \\
\midrule
Rater 1 & 0.750 & 0.762 & 0.30 & 0.558 \\
Rater 2 & 0.750 & 0.750 & 0.25 & 0.495 \\
\bottomrule
\end{tabular}
\end{minipage}
\end{table}

\smallskip
\noindent\textbf{Turing Test 2: Real versus synthetic discrimination with confidence.}
Radiologists classified each image as real or synthetic and assigned a five-point confidence score. As shown in Table~\ref{tab:turing} (b), both raters achieved moderate discrimination, indicating that some synthetic cues remain detectable. However, a substantial fraction of synthetic images were labeled as real, supporting their visual plausibility. Confidence was only weakly correlated with correctness, suggesting limited calibration in this setting.

\section{Conclusion}
Critical RFO detection represents a clinically important yet data-scarce regime, where positive samples are rare and diagnostic cues are subtle and highly localized, challenging both conventional supervised learning and end-to-end synthesis.
We introduced SurgRFO, a two-stage synthesis framework that separates surgical background modeling from localized RFO evidence generation. By combining a background diffusion generator with a lightweight local sampler through controlled fusion, SurgRFO produces high-quality RFO-positive radiographs for augmentation. We also released the Intraoperative RFO Bench, the first public benchmark dedicated to critical RFOs with expert annotations.
On this benchmark, SurgRFO-augmented training yielded consistent improvements across multiple detector families, including under cross-institutional evaluation. A blinded radiologist study further supported the clinical plausibility of synthetic images, with comparable quality ratings to real radiographs and only moderate discriminability.
These results demonstrate that structured synthesis can expand exposure to rare critical events and enhance detection sensitivity, while emphasizing the need for continued clinician-centered and workflow-aware validation prior to deployment.

%
%
%
%
\newpage
\bibliographystyle{splncs04}
\bibliography{mybibliography}

\end{document}